\documentclass[prd,twocolumn,altaffilletter,showpacs,nofootinbib]{revtex4}
\usepackage[dvips]{graphicx}
\usepackage{amsmath}

\newcommand{\be}{\begin{equation}}
\newcommand{\ee}{\end{equation}}
\newcommand{\bea}{\begin{eqnarray}}
\newcommand{\eea}{\end{eqnarray}}
\newcommand{\vphi}{\varphi}

\begin{document}

\title{Can noncommutative effects account for the present speed up of the cosmic expansion?}

\author{Octavio Obregon}\email{octavio@fisica.ugto.mx}\affiliation{Divisi\'on de Ciencias e Ingenier\'ia de la Universidad de Guanajuato, A.P. 150, 37150, Le\'on, Guanajuato, M\'exico.}

\author{Israel Quiros}\email{iquiros@fisica.ugto.mx}\affiliation{Divisi\'on de Ciencias e Ingenier\'ia de la Universidad de Guanajuato, A.P. 150, 37150, Le\'on, Guanajuato, M\'exico.}

\date{\today}

\begin{abstract}
In this paper we investigate to which extent noncommutativity, a intrinsically quantum property, may influence the Friedmann-Robertson-Walker cosmological dynamics at late times/large scales. To our purpose it will be enough to explore the asymptotic properties of the cosmological model in the phase space. Our recipe to build noncommutativity into our model is based in the approach of reference [Phys. Rev. Lett. {\bf 88} (2002) 161301], and can be summarized in the following steps: i) the Hamiltonian is derived from the Einstein-Hilbert action (plus a self-interacting scalar field action) for a Friedmann-Robertson-Walker spacetime with flat spatial sections, ii) canonical quantization recipe is applied, i. e., the minisuperspace variables are promoted to operators, and the WDW equation is written in terms of these variables, iii) noncommutativity in the minisuperspace is achieved through the replacement of the standard product of functions by the Moyal star product in the WDW equation, and, finally, iv) semi-classical cosmological equations are obtained by means of the WKB approximation applied to the (equivalent) modified Hamilton-Jacobi equation. We demonstrate, indeed, that noncommutative effects of the kind considered here, can be those responsible for the present speed up of the cosmic expansion.\end{abstract}

\maketitle

\section{Introduction}\label{intro}

Noncommutativity of spacetime coordinates -- an old idea dated back to 1947 \cite{snyder} -- is the simplest expected modification to quantum field theory. Noncommutativity -- the central mathematical concept in quantum mechanics -- expresses uncertainty in the simultaneous measurement of any pair of conjugate variables, such as position and momentum. In the presence of a strong magnetic field noncommutativity arises \cite{jackiw}, even in a classical context. More recently, noncommutativity has received increased interest in connection with developments in string theory. Attempts to connect M(atrix)-string theory to cosmology on the brane \cite{quevedo} have shown that noncommutativity arises in the former theory.

There are several approaches in the literature to build noncommutativity into field theories. One of the authors and collaborators have explored several of these approaches \cite{prl-basic, topological-nc-gravity, self-dual-nc-gravity, twisted-nc-gravity, compean-nc-cosmology, nc-lambda, mini-superspace-frw, gss, pimentel-obregon}. In some of these formalisms the assumption of noncommutativity among the spacetime coordinates has a consequence that the fields present do not commute themselves. This is the particular case of the Seiberg-Witten map \cite{seiberg-witten} and its generalization by Wess and collaborators \cite{new-ref}, in which noncommutative fields are obtained as an (infinite) expansion of the usual commutative fields in the noncommutative parameter. Space-time noncommutativity has as a consequence a different product of functions (the Moyal/star product), which induces noncommutativity among the fields.

In cosmological settings an already well explored way to include the effects of noncommutativity is given in reference \cite{prl-basic}, where deformation of spacetime itself is replaced by noncommutativity in minisuperspace instead (see also references \cite{citas-prl,citas-recientes-prl} where similar approaches are applied).\footnote{For further consideration of the approach of \cite{prl-basic} in the context of extra-dimensional cosmology see, for instance, Ref. \cite{jabbari} and references therein.} The latter proposal is inspired in various related results in the literature, among which we can cite the above mentioned case of the Seiberg-Witten map \cite{seiberg-witten,chamseddine,self-dual-nc-gravity}. In the case of quantum cosmology, the minisuperspace variables play the role of the coordinates in configuration space.\footnote{See, for instance, references \cite{hartle-hawking,halliwell}, for an alternative path integral approach to quantum cosmology.} Thus, as stated in \cite{prl-basic}, it seems reasonable to propose a kind of noncommutativity among these specific gravitational variables. The noncommutative proposal there is formulated in terms of models with a finite number of degrees of freedom, where the Wheeler-DeWitt (WDW) equation resembles a Klein-Gordon equation, this time in terms of the minisuperspace variables. One can then apply the same procedure as in noncommutative quantum mechanics to end up with a noncommutative version of quantum cosmology \cite{schrodinger}.

In the present paper we shall investigate the possible effect of noncommutativity in the late-time/large scale cosmological dynamics. In reference \cite{similar-dilaton} the effects of noncommutativity were investigated within a dilatonic cosmological model for an exponential dilaton potential. The existence of such noncommutativity results in a deformed Poisson algebra between the minisuperspace variables and their conjugated momenta. The authors found that noncommutativity modifies the cosmic dynamics at late times. Their result relies, however, on the study of an exact solution of the field equations that, in the commutative case, results in decelerating pace of expansion both at early and late times. It is clear that the latter is just a particular solution of the cosmological equations that can be attained only after a very careful arrangement of the initial conditions. Actually, as it was clearly and strictly shown in reference \cite{clw} -- through the use of the standard tools of the (linear) dynamical systems analysis -- the generic cosmological evolution at late times, in the case of a dilaton exponential potential-driven dynamics, depending on the region in the parameter space, can be governed either by an inflationary scalar field (SF)-dominated attractor solution, or by a scaling (also attractor) solution where neither the SF nor the background fluid dominate. In the latter case the ratio of the energy densities of the components of the cosmic mixture is a constant. As in the former case, at late times the expansion may be inflating even in this case. 

The results of the study in Ref. \cite{similar-dilaton}, although relevant in what respects the question put forward: can noncommutativity affect the dynamics of the universe in the large cosmological scales?, are of limited importance in what regards their reach. Actually, their results can not be generic since these are based on a particular exact solution.\footnote{In other words, the given solution may be unstable and very unlikely to arise in a real physical context.} In the present paper we aim at looking for a generic answer by the use of the standard tools of the dynamical systems analysis in the way the authors of Ref. \cite{clw} did. The basic steps of the formalism we use to build noncommutativity into our cosmological setting -- as we already said it is based on the approach of reference \cite{prl-basic} -- are the following: i) starting from the Einstein-Hilbert action (plus a self-interacting scalar field action), the Hamiltonian is derived for a Friedmann-Robertson-Walker spacetime with flat spatial sections, ii) canonical quantization recipe is applied: the minisuperspace variables are promoted to operators and the WDW equation is written in terms of these variables, iii) noncommutativity in the minisuperspace is achieved through the replacement of the standard product of functions by the Moyal star product, and iv) semi-classical cosmological equations are obtained by means of the Wentzel-Kramers-Brillouin (WKB) approximation,\footnote{Also known as Jeffreys-Wentzel-Kramers-Brillouin (JWKB) approximation \cite{wkb-classics}.} applied to the (equivalent) modified Hamilton-Jacobi equation \cite{wkb-method}. After this we choose adequate phase space variables and write the modified cosmological equations in the form of an autonomous system of ordinary differential equations (ODE). Then we follow the steps of Ref. \cite{clw}.

In the present investigation we do not restrict our study to exponential potential only, but include a most general situation when a cosh-like potential is also considered. Our results will indicate -- in as much strict terms as in \cite{clw} -- that the noncommutative effects not only affect he early-times dynamics, as expected due to their quantum nature, but also the cosmic dynamics at late times. 

Before going any further, we want to mention the main ''drawbacks'' of our approach: i) we will be considering noncommutativity in minisuperspace rather than in spacetime itself, and, ii) following the approach of Ref. \cite{prl-basic} we will be missing noncommutativity among the momenta conjugated of the minisuperspace variables. The former drawback is not as worrying as the latter one, since, although we are not dealing with direct spacetime noncommutativity, the kind of noncommutativity in minisuperspace we will be considering, is expected to be a derived consequence of the former \cite{prl-basic, seiberg-witten, topological-nc-gravity, self-dual-nc-gravity, twisted-nc-gravity, compean-nc-cosmology, nc-lambda, mini-superspace-frw, gss, pimentel-obregon, new-ref}. Although the latter drawback is of more concern, however, our approach here can be considered as a first step towards a more complete picture where noncommutativity of the conjugated momenta is also taken into account. 

We want to make emphasis in the fact that, given the above ''drawbacks'', and the ''freedom'' to choose minisuperspace coordinates in order to build noncommutativity into quantum cosmology \footnote{Recall that, even if one follows the more standard assumption of noncommutativity among spacetime coordinates, one would not obtain the same results assuming noncommutativity, e. g., among cartesian or spherical coordinates.} -- a theory yet lacking a strict formulation --, the cosmological model studied here can be considered, at most, as a useful toy model to qualitatively study possible cosmological consequences of noncommutative effects. In this sense, this is just a first attempt at the latter goal, and, consequently, in the present paper we shall be concerned mainly with the cosmological asymptotics and no attention will be paid to observational testing, a subject that deserves an independent publication. 

The paper has been organized in the following manner. In the next section \ref{setup} the basic mathematics of the model, as well as of the approach undertaken here, are exposed. We pay special attention to the following aspects: i) the canonical quantization procedure for minisuperspace variables, ii) introduction of noncommutativity through the Moyal star product, and, iii) derivation of the (semi-)classical cosmological equations by means of the WKB approximation. The main section \ref{dsystems} is devoted to the study of the asymptotic structure of the modified (semi-classical) cosmological model, through the use of the dynamical systems tools. The exponential and cosh type of potentials are studied separately. Figures with the relevant phase space portraits are supplied. The results obtained are discussed in detail in section \ref{discussion}. In section \ref{conclusion} the most important results of the paper are summarized and their reach commented on. We have added an appendix (section \ref{appendix}) where we discuss a possible generalization of our results to include the presence of cold dark matter (CDM) in general instead of a scalar field fluid, through the exploration of a simple toy model. Here we use units in which $\kappa^2=8\pi G=1$.

\section{Set up and basic equations}\label{setup}

As already mentioned in the introduction, here we follow the approach put forward for the first time in reference \cite{prl-basic}. This time, however, we consider a homogeneous and isotropic Friedmann-Robertson-Walker (FRW) spacetime with flat spatial sections. As a source of the Einstein's equations we choose a self-interacting scalar field $\vphi$. We will consider self-interaction potentials of the exponential and cosh-type.

\subsection{Hamiltonian approach}

To start with, we write the -- general relativity (GR) -- Einstein-Hilbert (EH) action:\footnote{For a pedagogical and readable explanation of the procedure we are about to expose see section 2 of Ref. \cite{halliwell}.} 

\be S_g=\frac{1}{2\kappa^2}\int d^4x\sqrt{-g}R,\label{gr-action}\ee where $\kappa^2\equiv 8\pi G$. If the homogeneous and isotropic FRW metric -- flat spatial sections -- is considered, then we can parametrize the metric as: $ds^2=-dt^2+e^{2\alpha(t)}\delta_{ij}dx^i dx^j,$ where $\alpha(t)$ is the time-dependent scale factor. Then, after integrating by parts, the above action can be re-written as, 

\be S_g=\frac{1}{\kappa^2}\int d^3x dt\; {\cal L}(\alpha,\dot\alpha),\label{gr-equiv-action}\ee where we have introduced the definition of the effective Lagrangean: ${\cal L}(\alpha,\dot\alpha)=-3 e^{3\alpha}\dot\alpha^2$. 

If, in addition to the standard EH (pure gravity) action $S_g$, one introduces also a self-interacting scalar field (SF) action:

\be S_\vphi=-\frac{1}{2\kappa^2}\int d^4x\sqrt{-g}\left[(\partial\vphi)^2+2V(\vphi)\right],\label{sf-action}\ee where $V(\vphi)$ is the self-interaction potential, while $(\partial\vphi)^2\equiv g^{\mu\nu}\partial_\mu\vphi\partial_\nu\vphi$, then, the total effective Lagrangean under the integral in the resulting action: $$S_{tot}=\frac{1}{\kappa^2}\int d^3x dt\; {\cal L}_{tot}(\alpha,\dot\alpha,\vphi,\dot\vphi),$$ can be written in the following form:

\be {\cal L}_{tot}= e^{3\alpha}\left[-3\dot\alpha^2+\frac{1}{2}\dot\vphi^2-V(\vphi)\right].\label{l-tot}\ee The Euler-Lagrange equations yield to the second Friedmann (also Raychaudhuri) equation, and to the Klein-Gordon equation, respectively: 

\be 2\ddot\alpha+3\dot\alpha^2=-\frac{1}{2}\dot\vphi^2+V(\vphi),\;\;\ddot\vphi+3\dot\alpha\dot\vphi=-V_{,\vphi},\label{feqs}\ee respectively. Here, and in what follows, $V_{,\vphi}\equiv\partial V/\partial\vphi$.

The next step is to go onto the equivalent Hamiltonian formulation. To this purpose we introduce the canonical momenta:

\be \pi_\alpha=\frac{\partial{\cal L}_{tot}}{\partial\dot\alpha}=-6e^{3\alpha}\dot\alpha,\;\;\pi_\vphi=\frac{\partial{\cal L}_{tot}}{\partial\dot\vphi}=e^{3\alpha}\dot\vphi,\label{canonical-m}\ee respectively. The standard relationship between the classical Hamiltonian ${\cal H}$ and the effective Lagrangean ${\cal L}_{tot}$: ${\cal H}(\alpha,\pi_\alpha,\vphi,\pi_\vphi)=\pi_\alpha\dot\alpha+\pi_\vphi\dot\vphi-{\cal L}_{tot}(\alpha,\dot\alpha,\vphi,\dot\vphi)$, yields to

\be {\cal H}=\frac{e^{-3\alpha}}{12}[-\pi_\alpha^2+6\pi_\vphi^2+12 e^{6\alpha} V(\vphi)].\label{h}\ee

\subsection{Mini-superspace variables and noncommutative WDW equation}

Here, following reference \cite{prl-basic}, noncommutativity is achieved by an appropriate deformation of the usual (commutative) algebra of minisuperspace variables $\alpha$, $\vphi$: 

\be [\alpha,\vphi]=0\;\rightarrow\;[\alpha',\vphi']=-i\theta,\label{commutator}\ee where the new minisuperspace variables $\alpha'$ and $\vphi'$ are noncommutative coordinates on a new (minisuperspace) base manifold \cite{mezincescu}, and the constant parameter $\theta$ encodes the noncommutative effects.\footnote{In what follows, for definiteness, we consider $\theta$ to be a non-negative quantity.} 

We think it is appropriate to make a comment on the physical motivations of the latter ansatz. Our assumption of noncommutativity among the expansion factor $\alpha'$ and the scalar field $\vphi'$ is based in the results of previous studies (see, for instance, references \cite{prl-basic, seiberg-witten, topological-nc-gravity, self-dual-nc-gravity, twisted-nc-gravity, compean-nc-cosmology, nc-lambda, mini-superspace-frw, gss, pimentel-obregon, new-ref}). In our model the parameter $\theta$ is expected to be a kind of effective (resulting) measure of noncommutativity amongst fields that are expanded in terms of the parameters related with noncommutativity of usual spacetime coordinates. The ansatz in noncommutative quantum cosmology is also inspired in noncommutative quantum mechanics, due to the fact that $\alpha$ ($\alpha'$) and $\vphi$ ($\vphi'$) play the role of ''coordinates'' in the minisuperspace, and appear as such in the WDW equation to study quantum cosmology \cite{ryan}.

Let us to promote the canonical momenta in (\ref{h}) to quantum operators ($\partial_\alpha\equiv\partial/\partial\alpha$, $\partial_\vphi\equiv\partial/\partial\vphi$): $\pi_\alpha\rightarrow\hat\pi_\alpha=-i\partial_\alpha,\;\;\pi_\vphi\rightarrow\hat\pi_\vphi=-i\partial_\vphi,$ i. e., 

\be {\cal H}\rightarrow\hat{\cal H}=\frac{e^{-3\alpha}}{12}\left[\partial_\alpha^2-6\partial_\vphi^2+12 e^{6\alpha} V(\vphi)\right].\label{quantum-h}\ee The minisuperspace WDW equation then reads:

\be \hat{\cal H}\Psi(\alpha,\vphi)=0,\label{wdw}\ee where $\Psi$ is the wave-function of the universe. 

Here we shall introduce the noncommutative effects through the Moyal star product of functions: 

\be \hat{\cal H}\Psi(\alpha,\vphi)=0\;\rightarrow\;\hat{\cal H}\star\Psi(\alpha,\vphi)=0.\label{moyal-star}\ee Since under the star operation the terms containing $\partial^2_\alpha$ and $\partial^2_\vphi$ are unchanged, the effects of the Moyal/star product are reflected in the WDW equation through a shift in the potential:

\be V(\alpha,\vphi)\star\Psi(\alpha,\vphi)=V(\alpha+\frac{\theta}{2}\hat\pi_\vphi,\vphi-\frac{\theta}{2}\hat\pi_\alpha)\Psi(\alpha,\vphi).\label{moyal}\ee  It has been shown that this is equivalent to performing the following change of variables \cite{mezincescu} (see also \cite{susskind}): 

\be \alpha'=\alpha+\frac{\theta}{2}\hat\pi_\vphi,\;\;\vphi'=\vphi-\frac{\theta}{2}\hat\pi_\alpha,\label{new-var}\ee where the involved field variables obey the corresponding commutation relationships in Eq. (\ref{commutator}).

To make the point clear: we can write the WDW equation modified through the replacement of the standard product of functions and operators by the Moyal/star product $\hat{\cal H}\star\Psi(\alpha,\vphi)=0$, which results in Eq. (\ref{nc-wdw}) (see below), or, equivalently, $\hat{\cal H}\Psi(\alpha',\vphi')=0$, formulated in terms of the noncommutative coordinates $\alpha'$, $\vphi'$, of the new base manifold. Here we adhere to the former point of view, i. e., we keep the ''old'' commutative minisuperspace variables $\alpha$, and $\vphi$, so that the noncommutativity will be encoded in the new terms containing the parameter $\theta$. The noncommutative counterpart of the WDW equation then reads:

\bea &&\hat{\cal H}_\theta\Psi(\alpha,\vphi)=0,\;\Rightarrow\;[\partial_\alpha^2-6\partial_\vphi^2+\hat U_\theta(\alpha,\vphi)]\Psi=0,\nonumber\\
&&\hat U_\theta(\alpha,\vphi)\equiv 12\;e^{(6\alpha-3i\theta\partial/\partial\vphi)}V(\vphi+i\frac{\theta}{2}\frac{\partial}{\partial\alpha}).\label{nc-wdw}\eea If we further Taylor expand in the small parameter $\theta$ -- this will entail assuming small $\theta\partial S(\alpha)/\partial\alpha$ and $\theta\partial S(\vphi)/\partial\vphi$ in the equivalent Hamilton-Jacobi formulation (see below) -- then, keeping up to linear terms:

\bea &&e^{(6\alpha-3i\theta\partial/\partial\vphi)}=e^{6\alpha}\left(1-3i\theta\frac{\partial}{\partial\vphi}+{\cal O}(\theta^2)\right),\nonumber\\
&&V(\vphi+\frac{i\theta}{2}\frac{\partial}{\partial\alpha})=V(\vphi)+\frac{i\theta}{2} V_{,\vphi}\frac{\partial}{\partial\alpha}+{\cal O}(\theta^2).\nonumber\eea 

Recalling that we demand the operators $\partial/\partial\alpha$ and $\partial/\partial\vphi$ to act exclusively on the wave function $\Psi$, it can be shown that, in the above considered linear approximation, the potential energy operator in equation (\ref{nc-wdw}), $\hat U_\theta$ can be written as follows: 

\be \hat U_\theta(\alpha,\vphi)=12\;e^{6\alpha}\left(V-3i\theta V\partial_\vphi+\frac{i\theta}{2} V_{,\vphi}\partial_\alpha\right).\label{taylor}\ee

\subsection{WKB approach}

To get a consistent description of the cosmological dynamics at any time we need a (semi) classical analog of the deformed WDW equation (\ref{nc-wdw}), (\ref{taylor}). Such a classical approximation to the obtained (modified) WDW equation may be based on the Hamilton-Jacobi formulation and the WKB approach \cite{wkb-method,wkb-classics}. Hence, following the lines of Ref. \cite{wkb-method} and references therein, lets assume separability of the wave-function in its arguments $\alpha$, $\vphi$: $$\Psi(\alpha,\vphi)\propto e^{iS(\alpha)+iS(\vphi)},$$ together with the following conditions on the derivatives of the $S$-functions: $$\left|\frac{\partial^2S(\alpha)}{\partial\alpha^2}\right|\ll\left(\frac{\partial S(\alpha)}{\partial\alpha}\right)^2,\;\;\left|\frac{\partial^2S(\vphi)}{\partial\vphi^2}\right|\ll\left(\frac{\partial S(\vphi)}{\partial\vphi}\right)^2,$$ then, by substituting back into (\ref{nc-wdw}), (\ref{taylor}), one obtains (recall that in the linear approximation considered here, one keeps terms up to ${\cal O}(\theta)$):

\bea &&\left(\frac{\partial S(\alpha)}{\partial\alpha}\right)^2-6\left(\frac{\partial S(\vphi)}{\partial\vphi}\right)^2=\nonumber\\&&\;\;\;\;\;\;\;12\;e^{6\alpha}\left(V+3\theta V\frac{\partial S(\vphi)}{\partial\vphi}-\frac{\theta}{2}V_{,\vphi}\frac{\partial S(\alpha)}{\partial\alpha}\right).\label{jacobi-eq-int}\eea Next, in order to get the classical equations, as it is customary in the standard WKB procedure, we make the following identifications:

\be \frac{\partial S(\alpha)}{\partial\alpha}=\pi_\alpha,\;\;\frac{\partial S(\vphi)}{\partial\vphi}=\pi_\vphi,\label{wkb-ident}\ee where $\pi_\alpha$ and $\pi_\vphi$, are just the canonical momenta given by Eq.(\ref{canonical-m}). Hence, after substituting back (\ref{wkb-ident}) into Eq. (\ref{jacobi-eq-int}) (taking into account the definitions of the canonical momenta in (Eq. (\ref{canonical-m})), one obtains the following modified Friedmann constraint:

\be 3\dot\alpha^2=\frac{1}{2}\dot\vphi^2+V+3\theta\;e^{3\alpha}\left(\dot\alpha V_{,\vphi}+\dot\vphi V\right).\label{wkb-friedmann-eq}\ee 

A dynamical equation to determine $\vphi$ can be derived coming back to the Hamiltonian constraints (\ref{nc-wdw}), (\ref{taylor}), which, in terms of the classical momenta $\pi_\alpha$, $\pi_\vphi$, result in: 

\bea &&{\cal H}_\theta=\frac{e^{-3\alpha}}{12}(-\pi_\alpha^2+6\pi_\vphi^2)+\nonumber\\
&&\;\;\;\;\;\;\;\;\;\;\;\;\;\;\;\;\;\;\;e^{3\alpha}\left(V+3\theta V\pi_\vphi-\frac{\theta}{2}V_{,\vphi}\pi_\alpha\right).\label{new-hamiltonian}\eea Here the noncommutative effect is encoded in the small parameter $\theta$. Then, by taking into account Eq. (\ref{canonical-m}), the canonical Hamilton's equation, $\dot\pi_\vphi=-\partial{\cal H}_\theta/\partial\vphi$, yields to the desired equation:

\be \ddot\vphi+3\dot\alpha\dot\vphi=-V_{,\vphi}-3\theta e^{3\alpha}(\dot\alpha V_{,\vphi\vphi}+\dot\vphi V_{,\vphi}).\label{wkb-kg-eq}\ee

\subsection{Cosmological equations}

Equations (\ref{wkb-friedmann-eq}) and (\ref{wkb-kg-eq}) form a closed system of differential equations. The Raychaudhuri (also, second Friedmann) equation can be derived by taking the time derivative of the Friedmann equation (\ref{wkb-friedmann-eq}) and substituting $\ddot\vphi$ from the modified Klein-Gordon equation (\ref{wkb-kg-eq}). We obtain 

\be 2\ddot\alpha+3\dot\alpha^2=-\frac{1}{2}\dot\vphi^2+V+3\theta e^{3\alpha}(\dot\alpha V_{,\vphi}+\dot\vphi V).\label{wkb-raycha-eq}\ee Equations (\ref{wkb-friedmann-eq}), (\ref{wkb-kg-eq}), and (\ref{wkb-raycha-eq}), are the modified GR equations of motion provided by the WKB approach, for the case where $\theta$ is a small quantity. 

A straightforward inspection of the modified cosmological equations show that the effect of the kind of noncommutativity explored in this paper is to modify the effective self-interaction potential of the scalar field. Actually, the second and third terms in the RHS of Eq. (\ref{wkb-raycha-eq}) can be joined together into the effective potential $V_{eff}$:

\be V_{eff}=V+3\theta\;e^{3\alpha}(\dot\alpha V_{,\vphi}+\dot\vphi V),\label{eff-pot}\ee which means that we can define the effective parametric energy density and pressure of the scalar fluid: $\rho^{eff}_\vphi=\dot\vphi^2/2+V_{eff}$, $p^{eff}_\vphi=\dot\vphi^2/2-V_{eff}$, respectively. Hence, the following set of "standard" field equations for a SF-sourced cosmology is obtained:

\bea &&3\dot\alpha^2=\rho^{eff}_\vphi,\;\;2\ddot\alpha+3\dot\alpha^2=-p^{eff}_\vphi,\nonumber\\
&&\dot{\rho}^{eff}_\vphi+3\dot\alpha(\rho^{eff}_\vphi+p^{eff}_\vphi)=0.\nonumber\eea 

An alternative interpretation is possible however. Actually, it is really tempting to identify the following effective energy density and pressure, respectively, of a ``noncommutative" (NC) vacuum fluid (see Ref. \cite{v-fluid} for a related model):

\be \rho_\theta=-p_\theta=3\theta\;e^{3\alpha}(\dot\alpha V_{,\vphi}+\dot\vphi V).\label{rho-p-theta}\ee After the above identification one may re-write the cosmological equations (\ref{wkb-friedmann-eq}), (\ref{wkb-kg-eq}), (\ref{wkb-raycha-eq}), in the following compact form (we also write the continuity equation for the effective noncommutative fluid):

\bea &&3\dot\alpha^2=\rho_\vphi+\rho_\theta,\;\;2\ddot\alpha+3\dot\alpha^2=-p_\vphi-p_\theta,\nonumber\\
&&\dot\rho_\vphi+3\dot\alpha(\rho_\vphi+p_\vphi)=-\dot\vphi\rho_\theta,\;\;\dot\rho_\theta=\dot\vphi\rho_\theta,\label{wkb-compact-eq}\eea where, as usual, $\rho_\vphi=\dot\vphi^2/2+V$, is the (parametric) energy density of the scalar field fluid, while $p_\vphi=\dot\vphi^2/2-V$, is its parametric pressure. 

Notice that, although the conservation equation is not obeyed by each separate component of the cosmic mixture, the mixture, as a whole, actually does obey the continuity equation: $\dot\rho_{tot}+3\dot\alpha(\rho_{tot}+p_{tot})=0$, where $\rho_{tot}=\rho_\vphi+\rho_\theta$, and $p_{tot}=p_\vphi+p_\theta$. It is evident that, written in the above suggestive form, the cosmic dynamics is governed by additional, non-gravitational interactions between the scalar field fluid and the effective NC fluid, through the source term $\dot\vphi\rho_\theta$. The dynamics of interacting cosmic fluids is a very interesting scenario to look for solutions to the cosmic coincidence problem \cite{zimdahl,ng-interactions,cqg2006}.

Another interesting feature of this alternative interpretation is that the effective NC fluid behaves just like a cosmological "constant" ($\rho_\theta=-p_\theta$), where, due to the source term $\dot\vphi\rho_\theta$ in the continuity equations, the cosmological constant is in fact a dynamical quantity. This means that the impact of noncommutativity on the cosmic expansion at late times may be non-vanishing.

In the next sections we shall explore the asymptotic properties of the above cosmological model, to look for generic features of the impact noncommutativity (of the kind considered here) might have on the large-scale cosmic dynamics.

\section{Dynamical systems}\label{dsystems}

Our task here will be to write the equations (\ref{wkb-friedmann-eq}), (\ref{wkb-kg-eq}), (\ref{wkb-raycha-eq}), in the form of an autonomous system of ordinary differential equations. To this end we have to choose appropriate phase space variables. In the present case our starting choice is the following:

\be x\equiv\frac{\dot\vphi}{\sqrt 6\dot\alpha},\;\;y\equiv\theta e^{3\alpha}\dot\alpha,\;\;z\equiv-\frac{V_{,\vphi}}{V}.\label{var}\ee Notice that $x$ amounts to the dimensionless kinetic energy of the scalar field squared, while $y$ encodes the noncommutative effects. The phase space variable $z$ is sensitive to the kind of self-interaction potential chosen. In what follows we restrict our study to non-negative $z$-s, since the negative sector can be obtained from the non-negative one by the transformation $\vphi\rightarrow -\vphi$. 

The Friedmann equation (\ref{wkb-friedmann-eq}) yields to the following constraint relating the dimensionless SF potential energy with the variables $x$, $y$, $z$:

\be \frac{V}{3\dot\alpha^2}=\frac{1-x^2}{1+3y(\sqrt 6 x-z)}.\label{constraint}\ee 

The autonomous system of ODE that can be obtained out of the cosmological equations (\ref{wkb-friedmann-eq}), (\ref{wkb-kg-eq}), (\ref{wkb-raycha-eq}), is the following:

\bea &&x'=-3x(1-x^2)+\nonumber\\
&&\;\;\;\;\;\;\;\;\;\;\;\;\;\;\;\;\sqrt\frac{3}{2}(1-x^2)\left[\frac{1+3y(\sqrt 6 x-z\Gamma)}{1+3y(\sqrt 6 x-z)}\right]z,\nonumber\\
&&y'=3y(1-x^2),\;\;z'=-\sqrt 6 x z^2(\Gamma-1),\label{asode}\eea where $\Gamma\equiv VV_{,\vphi\vphi}/V_{,\vphi}^2$, and the tilde denotes derivative with respect to the time-ordering variable $\alpha=\ln a$ (basically the number of e-foldings). 

The above system of ODE is a closed system of ordinary differential equations only if the parameter $\Gamma$ can be written as a function of the variable $z$; $\Gamma=\Gamma(z)$. Indeed, for a large class of self-interaction potentials which are very popular in the cosmological context, $\Gamma$ can be written in the form of a polynomial in $z$. 

The phase space relevant for this problem depends on the particular kind of potential considered, but, in general it can be defined as:

\be \Phi=\{(x,y,z)|y\geq 0,\;z\geq 0...\}.\label{phase-space-general}\ee The dots mean additional constraints on the phase space variables, coming from the particular functional form of the self-interaction potential under consideration. We choose $y\geq 0$ since we are interested in expanding cosmologies only ($\dot\alpha>0$, recall that we have chosen $\theta$ to be a non-negative quantity). 

Several magnitudes of cosmological relevance are the following. The deceleration parameter ($q=-(1+\ddot\alpha/\dot\alpha^2)$):

\be q=-1+3x^2,\label{deceleration-p}\ee and the dimensionless energy densities of the NC fluid and of the SF:

\be \Omega_\theta\equiv\frac{\rho_\theta}{3\dot\alpha^2}=\frac{3(1-x^2)(\sqrt 6 x-z)y}{1+3y(\sqrt 6 x-z)},\;\;\Omega_\vphi=1-\Omega_\theta,\label{density-p}\ee respectively. Recall that, since the effective NC fluid behaves like vacuum ($p_\theta=-\rho_\theta$), then the effective equation of state (EOS) parameter, $\omega_\theta=-1$, always.

In general, for potentials which vanish at the minimum -- usually associated with relevant late-time behavior -- the phase space variable $z$ is undefined, and one have to choose an alternative variable. One possible choice is the following:

\be \bar z=\frac{1}{z+1}\;\;\Rightarrow\;\;z=\frac{1-\bar z}{\bar z},\label{z-bar}\ee where, since $z$ is a non-negative quantity, then $0\leq\bar z\leq 1$. Besides, the resulting phase space spanned by the variables $x$, $y$, $\bar z$ is noncompact only in the $y$-direction, so that several critical points might scape our study. A way out is to transform also the variable $y$: $$\bar y=\frac{1}{y+1}\;\;\Rightarrow\;\;y=\frac{1-\bar y}{\bar y}.$$ After the above transformations the phase space spanned by $x$, $\bar y$, $\bar z$, is a compact space, which means that all of the existing equilibrium points are confined to a bounded region. 

In terms of $x$, $\bar y$, $\bar z$, the autonomous system of ODE (\ref{asode}) transforms into:

\bea &&x'=-3x(1-x^2)+\sqrt\frac{3}{2}(1-x^2)\left(\frac{1-\bar z}{\bar z}\right)\times\nonumber\\
&&\times\left\{\frac{\bar y\bar z+3(1-\bar y)[\sqrt 6 x\bar z-(1-\bar z)\Gamma(\bar z)]}{\bar y\bar z+3(1-\bar y)[\sqrt 6 x\bar z-(1-\bar z)]}\right\},\nonumber\\
&&\bar y'=-3\bar y(1-\bar y)(1-x^2),\nonumber\\
&&\bar z'=\sqrt 6 x (1-\bar z)^2[\Gamma(\bar z)-1].\label{asode-bar}\eea 

As already said, the phase space is now a compact 3-dimensional space spanned by the coordinates $x$, $\bar y$, $\bar z$, and can be defined in the following way:

\be \Phi=\{(x,\bar y,\bar z)|-1\leq x\leq 1,\;0\leq\bar y\leq 1,\;0\leq\bar z\leq1\}.\label{compact-phac-spac}\ee Recall that the ranges $\bar y>1$ and $\bar y<0$, are not being considered since we are interested in describing the dynamics of expansion (contracting phases are not of interest to us). 

Such relevant cosmological parameters as the SF and NC dimensionless energy density parameters, together with the scalar field EOS parameter, can be given in terms of the new variables as follows: $$\Omega_\theta=\frac{3(1-x^2)(1-\bar y)[(\sqrt 6 x+1)\bar z-1]}{\bar y\bar z+3(1-\bar y)[(\sqrt 6 x+1)\bar z-1]},\;\;\Omega_\vphi=1-\Omega_\theta,$$ and $$\omega_\vphi=1+\frac{2(x^2-1)\bar y\bar z}{\bar y\bar z+3x^2(1-\bar y)[(\sqrt 6 x+1)\bar z-1]},$$ respectively. Other relevant magnitudes remain unchanged: $q=-1+3x^2$, $\omega_\theta=-1$.

\begin{figure}[t!]
\begin{center}
\includegraphics[width=4.1cm,height=4cm]{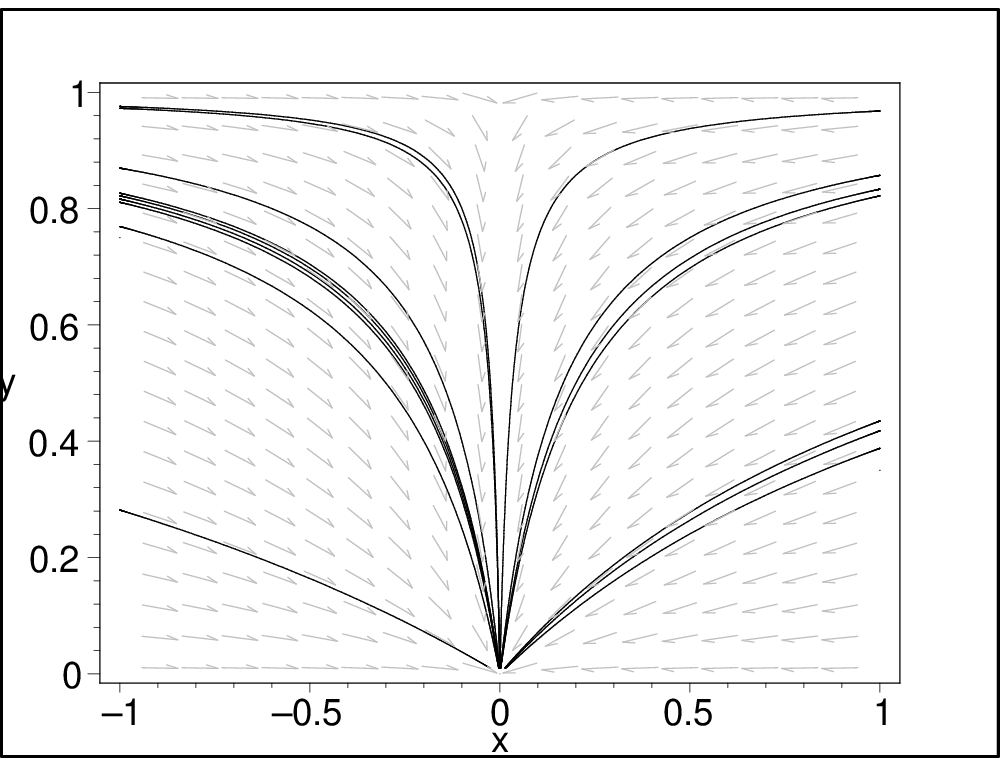}
\includegraphics[width=4.1cm,height=4cm]{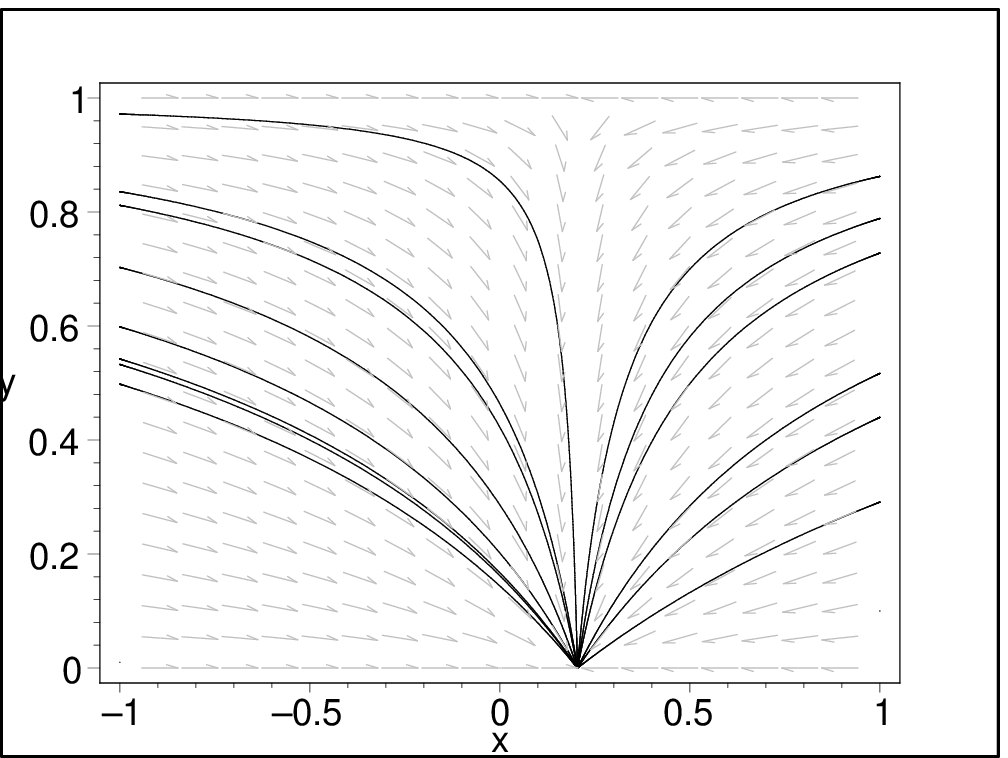}
\includegraphics[width=5cm,height=4cm]{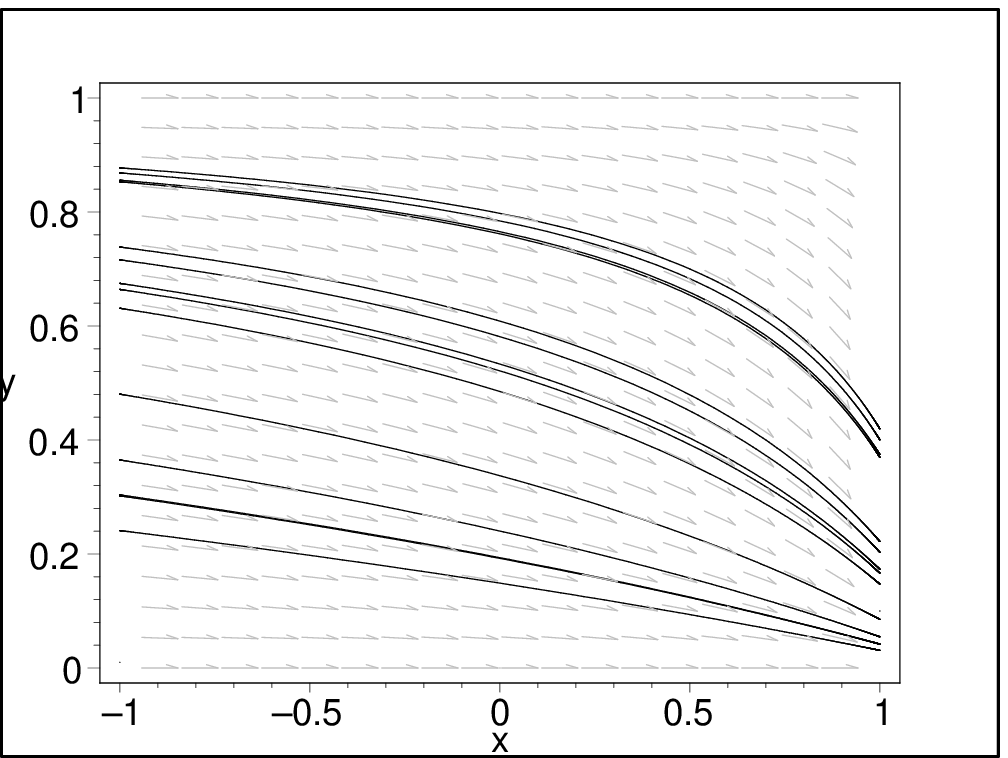}
\end{center}
\caption{Phase portraits for the exponential potential case (subsection \ref{exp}) for different choices of the exponential parameter $\lambda$: i) $\lambda=0$ (top left), ii) $\lambda=0.5$ (top right), and iii) $\lambda=3$ (bottom). In this, as well as in the subsequent figures, the $x$, $y$ (and $z$) axes span the compact phase space variables $x$, $\bar y$ (and $\bar z$) respectively. Several orbits corresponding to different sets of initial conditions are shown. In each case the initial conditions on the ``noncommutative'' coordinate $\bar y$ correspond to either $\bar y(0)=0.75$ (upper groups of orbits), or to $\bar y(0)=0.25$ (lower groups of orbits) only. The attractor structure of the scalar field/noncommutative fluid scaling solution (critical point $P_{\vphi/\theta}$) when $\lambda^2<6$ (upper panels), is evident.} \label{fig1}
\end{figure}

\subsection{Exponential potential}\label{exp}

For the particular case when $z=\lambda\Rightarrow\bar z=1/(\lambda+1)$, is a constant: $V(\vphi)=V_0\exp(-\lambda\vphi)\;\Rightarrow\;\Gamma=1$. In this case the system of ODE (\ref{asode-bar}) appreciably simplifies, and instead of a 3-dimensional phase space, one have a 2-dimensional one, spanned by the variables $x$, $\bar y$:

\bea &&x'=-\sqrt\frac{3}{2}(1-x^2)(\sqrt 6 x-\lambda),\nonumber\\
&&\bar y'=-3\bar y(1-\bar y)(1-x^2).\label{exp-asode}\eea We have, also, that $q=-1+3x^2$ (as before), and $$\Omega_\theta=\frac{3(1-x^2)(\sqrt 6 x-\lambda)(1-\bar y)}{\bar y+3(\sqrt 6 x-\lambda)(1-\bar y)}.$$ 

The main qualitative properties of the phase space spanned by the variables $x$, $\bar y$, $\bar z$, for this case, are summarized below.

\subsubsection{Critical points: commutative (GR) limit}

In the commutative GR limit ($y=0\;\Rightarrow\;\bar y=1$), the phase space is a linear space, and the critical points are $x=\pm 1$ and $x=\lambda/\sqrt 6$. Perturbing (linearly) in the neighborhood of these points one obtains: $$\epsilon'_\pm=6\left(1\mp\frac{\lambda}{\sqrt 6}\right)\epsilon,\;\;\epsilon'=3\left(\frac{\lambda^2}{6}-1\right)\epsilon,$$ respectively, where $\epsilon$ is the small perturbation. Hence, the perturbation evolves according to $\epsilon_\pm(\alpha)=\epsilon_0\exp((6\mp\sqrt 6\lambda)\alpha)$, for the points $x=\pm 1$, and $\epsilon(\alpha)=\epsilon_0\exp((\lambda^2-6)\alpha/2)$, for $x=\lambda/\sqrt 6$, respectively. This means that, while the SF kinetic energy-dominated solution corresponding to the choice $x=1$ (also stiff-fluid solution), can be stable for $\lambda>\sqrt 6$, the one corresponding to $x=-1$ is always unstable. For $\lambda^2<6$ the SF kinetic/potential energy-scaling solution $x=\lambda/\sqrt 6$ is stable, otherwise it is also an unstable critical point.

\subsubsection{Critical points: general case (includes noncommutativity)}

In the general case (including noncommutativity), the critical points, $P_i=(x_i,\bar y_i)$, of the above autonomous system of ODE are: 

\begin{itemize}

\item{SF kinetic energy-dominated solution:} $P_K^\pm=(\pm 1,\bar y)$. The cosmological parameters, in this case, are: $$\Omega_\theta=0,\;\;\Omega_\vphi=1,\;\;q=2.$$ Since, in both cases, there is one vanishing eigenvalue of the corresponding linearization matrices: i) $\lambda_1=0$, $\lambda_2=6-\sqrt 6\lambda$, for $P_K^+$, ii) $\lambda_1=0$, $\lambda_2=6+\sqrt 6\lambda$, for $P_K^-$, these critical points are non-hyperbolic.

\item{SF scaling solution:} $P_{K/V}=(\lambda/\sqrt 6,1)$. The dimensionless density parameters, and the deceleration parameter are equal to: $$\Omega_\theta=0,\;\;\Omega_\vphi=1,\;\;q=-1+\frac{\lambda^2}{2}.$$ The eigenvalues of the linearization around this point are: $\lambda_1=3-\lambda^2/2$, $\lambda_2=-3+\lambda^2/2$, so that it is a saddle equilibrium point.

\item{SF/NC fluid-scaling solution:} $P_{\vphi/\theta}=(\lambda/\sqrt 6,0)$. The relevant cosmological parameters are: $$\Omega_\theta=1-\frac{\lambda^2}{6},\;\;\Omega_\vphi=\frac{\lambda^2}{6},\;\;q=-1+\frac{\lambda^2}{2},$$ while the eigenvalues of the linearization matrix are: $\lambda_1=\lambda_2=-3+\lambda^2/6$. For $\lambda^2<6$ this is the future attractor in the phase space.

\end{itemize}

Summarizing the results of this subsection: i) the noncommutative effects modify the early-times cosmic dynamics by erasing any past attractor (see the figure \ref{fig1}), ii) these effects also modify the late-times dynamics by replacing the two possible future attractors according to the standard GR-limit (either the stiff-fluid solution, or the SF kinetic/potential energy-scaling solution) by the SF/NC-scaling solution (critical point $P_{\vphi/\theta}$).

\begin{table*}[tbp]\caption[crit]{Critical points and their properties for the cosh-like potential $V(\vphi)=V_0 \left[\cosh(\lambda\vphi)-1\right]^p$: the commutative GR limit. The eigenvalues of the corresponding linearization matrices are shown in the right-hand columns.}
\begin{tabular}{@{\hspace{4pt}}c@{\hspace{14pt}}c@{\hspace{14pt}}c@{\hspace{14pt}}c@{\hspace{14pt}}c@{\hspace{14pt}}c@{\hspace{14pt}}c@{\hspace{14pt}}c}
\hline\hline\\[-0.3cm]
C. Points &$x$&$\bar z$& Existence & $\omega_\vphi$& $q$ & $\lambda_1$& $\lambda_2$ \\[0.1cm]
\hline\\[-0.2cm]
$P^+_K$ & $1$ & $\frac{1}{1+p\lambda}$ & Always & $1$ & $2$ & $6-\sqrt 6 p\lambda$ & $\sqrt 6\lambda$ \\[0.2cm]
$P^-_K$ & $-1$ & $\frac{1}{1+p\lambda}$ & " & $1$ & $2$ & $6+\sqrt 6 p\lambda$ & $-\sqrt 6\lambda$ \\[0.2cm]
$P_{K/V}$ & $\frac{p\lambda}{\sqrt 6}$ & $\frac{1}{1+p\lambda}$ & '' & $-1+\frac{p^2\lambda^2}{3}$ & $-1+\frac{p^2\lambda^2}{2}$ & $-3+\frac{p^2\lambda^2}{2}$ & $p\lambda^2$ \\[0.2cm]
$P_V$ & $0$ & $1$ & " & $-1$ & $-1$ & $-\frac{3}{2}\left(1+\sqrt{1-\frac{2}{3}p\lambda^2}\right)$ & $-\frac{3}{2}\left(1-\sqrt{1-\frac{2}{3}p\lambda^2}\right)$ \\[0.2cm]
\hline \hline
\end{tabular}\label{tab1}
\end{table*}

\begin{table*}[tbp]\caption[crit]{The critical points of the autonomous system (\ref{asode-bar}) for the cosh-like potential $V(\vphi)=V_0 \left[\cosh(\lambda\vphi)-1\right]^p$: general case (includes noncommutativity). Here we have defined $r=(2p-1+\sqrt{p^2\lambda^2(1-2p)})/(2p-1+p^2\lambda^2)$.}
\begin{tabular}{@{\hspace{4pt}}c@{\hspace{14pt}}c@{\hspace{14pt}}c@{\hspace{14pt}}c@{\hspace{14pt}}c@{\hspace{14pt}}c@{\hspace{14pt}}c@{\hspace{14pt}}c@{\hspace{14pt}}c}
\hline\hline\\[-0.3cm]
C. Points &$x$&$\bar y$&$\bar z$&Existence& $\Omega_\vphi$& $\Omega_\theta$ & $\omega_\vphi$ & $q$\\[0.1cm]
\hline\\[-0.2cm]
$P^+_K$ & $1$ & $\bar y$ & $\frac{1}{1+p\lambda}$ & Always & $1$ & $0$ & $1$ & $2$ \\[0.2cm]
$P^-_K$ & $-1$ & $\bar y$ & $\frac{1}{1+p\lambda}$ & '' & $1$ & $0$ & $1$ & $2$ \\[0.2cm]
$P_{K/V}$ & $\frac{p\lambda}{\sqrt 6}$ & $1$ & $\frac{1}{1+p\lambda}$ & '' & $1$ & $0$ & $-1+\frac{p^2\lambda^2}{3}$ & $-1+\frac{p^2\lambda^2}{2}$ \\[0.2cm]
$P_V$ & $0$ & $1$ & $1$ & '' & $1$ & $0$ & $-1$ & $-1$ \\[0.2cm]
$P^*_V$ & $0$ & $1$ & $0$ & '' & Undefined & Undefined & $-1$ & $-1$ \\[0.2cm]
$P_{NC}$ & $0$ & $0$ & $r$ & $p\leq 1/2$ & $0$ & $1$& $-1$ & $-1$ \\[0.2cm]
\hline \hline
\end{tabular}\label{tab2}
\end{table*}

\begin{table*}[tbp]\caption[crit]{Eigenvalues of the linearization matrices corresponding to the critical points in Tab.\ref{tab2}. Here we have defined: $m\equiv 3\sqrt{1-(\sqrt{1-2p}-p^2\lambda^2)^2/6p^2}$.}
\begin{tabular}{@{\hspace{4pt}}c@{\hspace{14pt}}c@{\hspace{14pt}}c@{\hspace{14pt}}c@{\hspace{14pt}}c@{\hspace{14pt}}c@{\hspace{14pt}}c}
\hline\hline\\[-0.3cm]
C. Points &$x$&$\bar y$&$\bar z$& $\lambda_1$ & $\lambda_2$& $\lambda_3$ \\[0.1cm]
\hline\\[-0.2cm]
$P^+_K$ & $1$ & $\bar y$ & $\frac{1}{1+p\lambda}$ & $0$ & $6-\sqrt 6 p\lambda$ & $\sqrt 6\lambda$ \\[0.2cm]
$P^-_K$ & $-1$ & $\bar y$ & $\frac{1}{1+p\lambda}$ & $0$ & $6+\sqrt 6 p\lambda$ & $-\sqrt 6\lambda$ \\[0.2cm]
$P_{K/V}$ & $\frac{p\lambda}{\sqrt 6}$ & $1$ & $\frac{1}{1\pm p\lambda}$ & $p\lambda^2$ & $-3+\frac{p^2\lambda^2}{2}$ & $3-\frac{p^2\lambda^2}{2}$ \\[0.2cm]
$P_V$ & $0$ & $1$ & $1$ & $3$ & $-3+m$ & $-3-m$ \\[0.2cm]
$P^*_V$ & $0$ & $1$ & $0$ & $3$ & $-3+m$ & $-3-m$ \\[0.2cm]
$P_{NC}$ & $0$ & $0$ & $r$ & $-3$ & $-3+\sqrt{9-3p\lambda^2}$& $-3-\sqrt{9-3p\lambda^2}$ \\[0.2cm]
\hline \hline
\end{tabular}\label{tab3}
\end{table*}

\subsection{Potentials of the cosh-like type}\label{cosh-like}

Here one is considering potentials of the following kind:

\be V(\vphi)=V_0\left[\cosh(\lambda\vphi)-1\right]^p,\label{cosh-pot}\ee so that $$z=\frac{p\lambda\sinh(\lambda\vphi)}{\cosh(\lambda\vphi)-1},\,\;\Gamma(z)=1-\frac{1}{2p}+\frac{p\lambda^2}{2z^2}.$$ Hence, in terms of the variable $\bar z$: $$\Gamma(\bar z)=\frac{(2p-1)(1-\bar z)^2+p^2\lambda^2\bar z^2}{2p(1-\bar z)^2},$$ besides; $$\Gamma(\bar z)-1=\frac{(p^2\lambda^2-1)\bar z^2+2\bar z-1}{2p(1-\bar z)^2}.$$ 

If we substitute back into (\ref{asode-bar}), and then we look for the equilibrium points of the resulting autonomous system of ODE, we obtain the results which we list below (see also the tables \ref{tab1}, \ref{tab2}, and \ref{tab3}).

\subsubsection{Critical points: commutative (GR) limit}\label{cosh-gr}

The commutative case corresponds to the choice $\bar y=1$, so that $y=0\;\Rightarrow\;\theta=0$ (the case $\dot\alpha=0$ is of no interest for cosmology). The corresponding critical points and their properties are summarized in table \ref{tab1}. Here we list these results. Recall that critical points are given by pairs $(x,\bar z)$.

\begin{figure}[t!]
\begin{center}
\includegraphics[width=7cm,height=5cm]{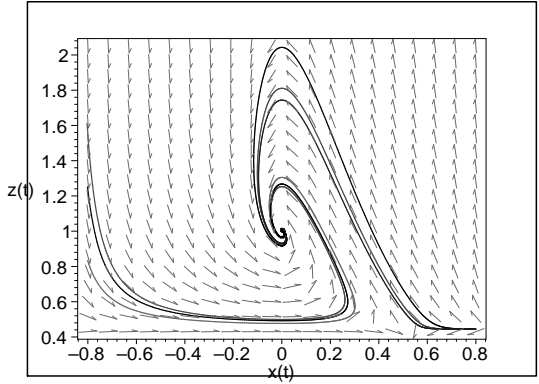}
\end{center}
\caption{Standard GR-limit for the cosh-like potential. The following values of the free parameters have been chosen: $\lambda=5$, $p=0.25$. Several orbits corresponding to different sets of initial conditions are drawn. The SF potential energy-dominated solution $P_V=(0,1)$ is the late-time attractor. It is seen that the orbits of the autonomous system of ODE spiral down to the future attractor, which is associated with coherent (damped) oscillations of the cosmological scalar field around the minimum of the potential.} \label{fig2}
\end{figure}

\begin{itemize}

\item Kinetic energy-dominated critical points: $$P^\pm_K=\left(\pm 1,\frac{1}{1+p\lambda}\right),$$ also known as stiff-fluid solution. For these points $\omega_\vphi=1$, and $q=2$, while the eigenvalues of the linearization matrix are $\lambda_1=6-\sqrt 6 p\lambda$, and $\lambda_2=\sqrt 6\lambda$, for $P^+_K$, while for $P^-_K$ these are: $\lambda_1=6+\sqrt 6 p\lambda$, $\lambda_2=-\sqrt 6\lambda$. Hence, for $p\lambda<\sqrt 6$, $P^+_K$ is the past attractor in the phase space, while, for $p\lambda>\sqrt 6$, it is a saddle critical point instead. The critical point $P^-_K$ is always a saddle point in the phase space.

\item SF kinetic/potential energy-scaling solution: $$P_{K/V}=\left(\frac{p\lambda}{\sqrt 6},\frac{1}{1+p\lambda}\right).$$ For this case the SF EOS and the deceleration parameters are given by $\omega_\vphi=-1+p^2\lambda^2/3$, $q=-1+p^2\lambda^2/2$, respectively. Since the eigenvalues of the linearization matrix for this case $\lambda_1=-3+p^2\lambda^2/2$, $\lambda_2=p\lambda^2$, then this critical point is a source (past attractor) whenever $p^2\lambda^2>6$, while for $p^2\lambda^2<6$ it is a saddle equilibrium point. For $p^2\lambda^2<2$ the corresponding cosmological solution describes accelerated expansion.

\item SF potential energy (V)-dominated solution: $P_V=(0,1)$. The scalar field fluid mimics a cosmological constant since $\omega_\vphi=-1$ and the peace of the cosmic expansion is accelerated ($q=-1$). This solution is always the future attractor since the real parts of the eigenvalues of the linearization matrix are both negative: $2\lambda_{1,2}=-3(1\pm\sqrt{1-2p\lambda^2/3})$. For $p\lambda^2>3/2$ the future attractor is a spiral equilibrium point (see the Fig. \ref{fig2}).

\end{itemize}

Summing up: within the framework of standard (commutative) GR-limit, the past attractor can be either the stiff-fluid solution if $p\lambda<\sqrt 6$ (the K/V energy-scaling solution and the conjugated stiff fluid solution are both saddle critical points), or the K/V energy-dominated solution if $p\lambda>\sqrt 6$. In the latter case the stiff fluid solution (and its indistinguishable conjugated, in this case) is always a saddle equilibrium point. The future (late-time) attractor -- a spiral point for $p\lambda^2>3/2$ -- is always the inflationary SF potential energy-dominated solution. These features are illustrated in the phase portrait in the figure \ref{fig2}.

\subsubsection{Critical points: general case}\label{cosh-general}

While in the GR-limit there are found four critical points, in the general case -- considering NC effects of the kind explored here -- there are found two additional critical points. Although all these critical points, $P_i=(x_i,\bar y_i,\bar z_i)$, are exposed in the table \ref{tab2} (the eigenvalues of the corresponding linearization matrices are shown in the table \ref{tab3}), here we list them and discuss their basic properties.

\begin{figure}[t!]
\begin{center}
\includegraphics[width=7cm,height=5cm]{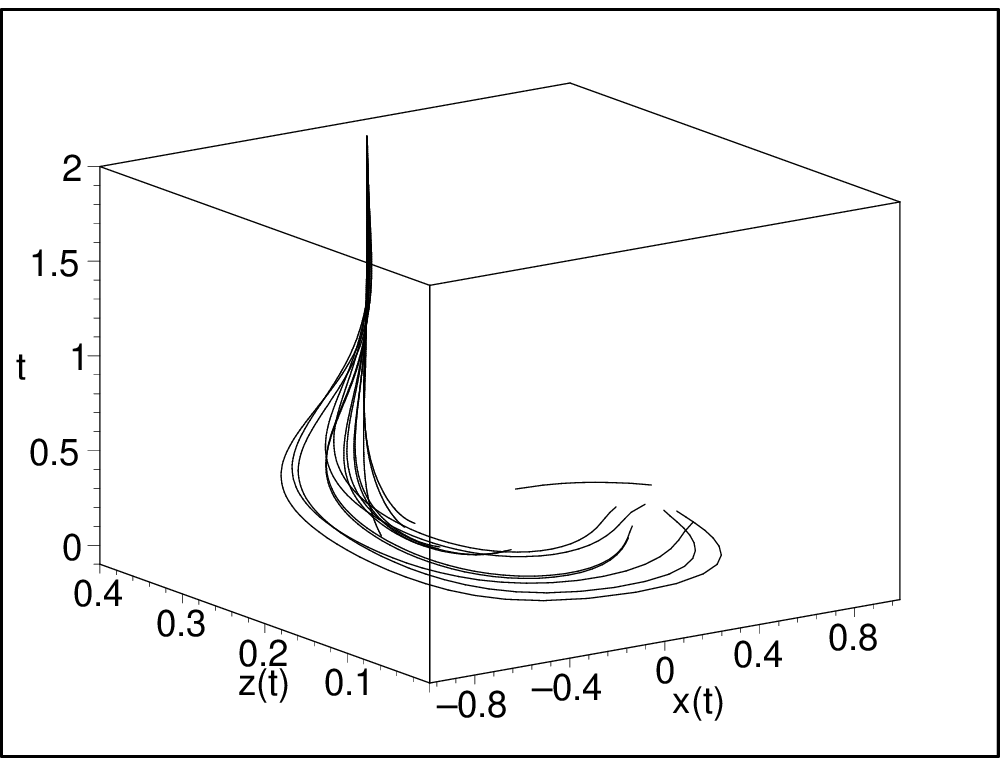}
\includegraphics[width=4.1cm,height=4cm]{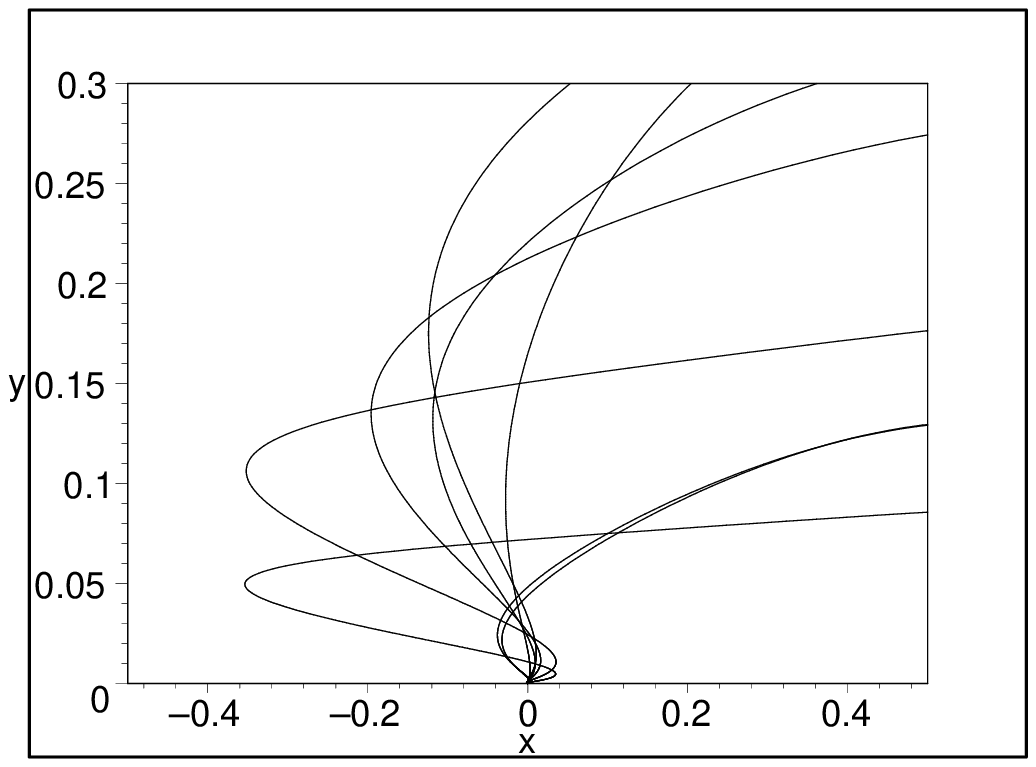}
\includegraphics[width=4.1cm,height=4cm]{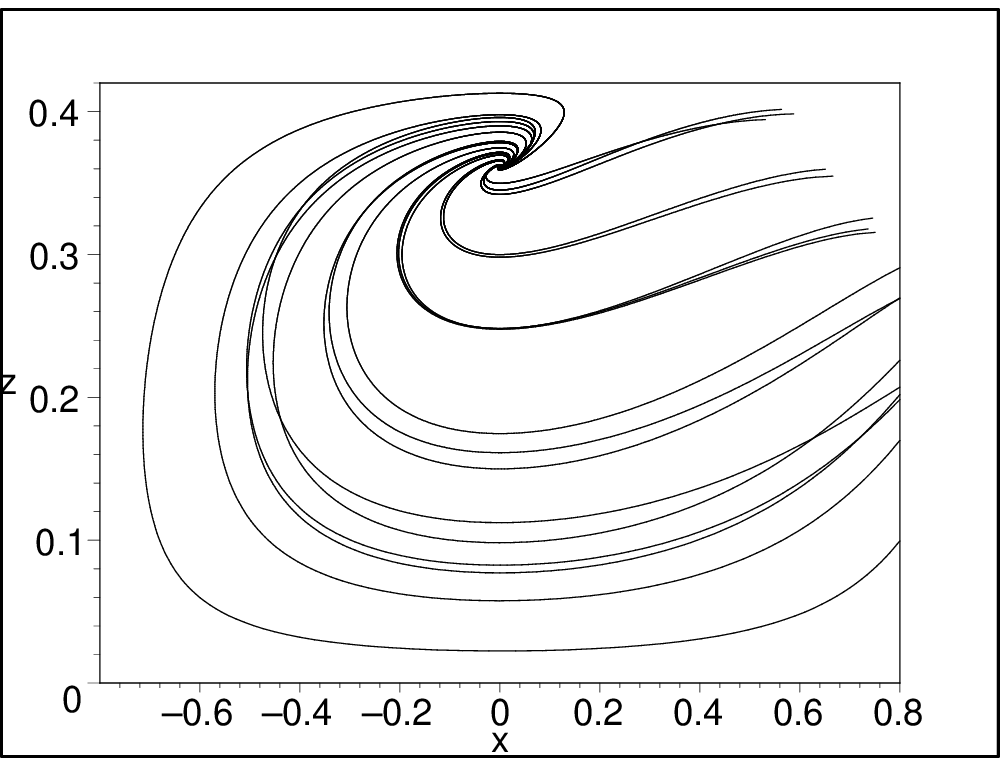}
\end{center}
\caption{General case for the cosh-like potential (includes noncommutative effects). The flux in time $\alpha$ of the corresponding autonomous system of ODE is shown for $\lambda=5$, $p=0.25$ (top). The projections of the phase space onto the different phase planes are also shown (bottom). While the existence of the future attractor -- critical point $P_{NC}=(0,0,0.36)$ -- corresponding to the solution dominated by the noncommutative effects (see the table \ref{tab2}), is evident, there are not found past attractors. Due to our choice of the free parameters above, $p\lambda^2>3$, the late-time attractor is a spiral critical point. For $p>1/2$ no future attractor can be found in the phase space neither.} \label{fig3}
\end{figure}

\begin{itemize}

\item Kinetic energy-dominated critical points (stiff fluid solution): $$P^\pm_K=\left(\pm 1,\bar y,\frac{1}{1+p\lambda}\right).$$ For these points $\Omega_\vphi=1$ ($\Omega_\theta=0$), $\omega_\vphi=1$, and $q=2$. For $P^+_K$ the eigenvalues of the linearization matrix are $\lambda_1=0$, $\lambda_2=6-\sqrt 6 p\lambda$, and $\lambda_3=\sqrt 6\lambda$, while, for $P^-_K$ these are: $\lambda_1=0$, $\lambda_2=6+\sqrt 6 p\lambda$, and $\lambda_3=-\sqrt 6\lambda$. In both cases, since one of the eigenvalues is vanishing, these are non-hyperbolic points and we miss part of the information we could obtain from the application of the standard tools of the (linear) dynamical systems analysis. However, for $P^+_K$, as long as $p\lambda>\sqrt 6$, this is a saddle equilibrium point since $\lambda_2$ and $\lambda_3$ are of opposite sign. In the case $P^-_K$, since for positive $p$ the eigenvalues $\lambda_2$ and $\lambda_3$ are of opposite sign, then we can say with certainty that this is always a saddle critical point. Additional information can be extracted from the inspection of the phase portrait (see Fig. \ref{fig3}). It can be corroborated that there are no past attractors there.

\item SF K/V energy-scaling solution: $$P_{K/V}=\left(\frac{p\lambda}{\sqrt 6},1,\frac{1}{1+p\lambda}\right).$$ The cosmological magnitudes of relevance are given by: $\Omega_\vphi=1$ ($\Omega_\theta=0$), $\omega_\vphi=-1+p^2\lambda^2/3$, $q=-1+p^2\lambda^2/2$. The eigenvalues of the linearization matrix for this case are: $\lambda_1=p\lambda^2$, $\lambda_2=-3+p^2\lambda^2/2$, $\lambda_3=3-p^2\lambda^2/2$. Since the eigenvalues $\lambda_2$ and $\lambda_3$ are always of opposite sign, then this critical point is always a saddle point in the phase space, unlike in the GR-limit, when this point can be also a past attractor.

\item SF V-dominated solution: $P_V=(0,1,1)$. The relevant parameters are $\Omega_\vphi=1$ ($\Omega_\theta=0$), $\omega_\vphi=-1$, and $q=-1$, while the eigenvalues of the linearization matrix are: $\lambda_1=3$, $\lambda_{2,3}=-3\pm m$ (the quantity $m$ is defined in the heading of Tab. \ref{tab3}). This is always a saddle equilibrium point in the phase space.

\item Conjugated SF V-dominated critical point: $P^*_V=(0,1,0)$. In this case both $\Omega_\vphi$ and $\Omega_\theta$ are undefined. The other quantities remain the same as in the former case, including the eigenvalues of the linearization matrix. This point can be associated with the minimum of the cosh-like potential (\ref{cosh-pot}). 

\item NC-dominated solution: $P_{NC}=(0,0,r)$, where the quantity $r$ has been defined in the heading of Tab. \ref{tab2}. This solution is dominated by the energy density of the effective noncommutative fluid $\Omega_\theta=1$ ($\Omega_\vphi=0$), and exist whenever $p\leq 1/2$. It is associated with accelerated expansion since $q=-1$. Whenever it exists, it is the late-time (future) attractor since the real parts of the eigenvalues of the corresponding linearization matrix are negative: $\lambda_1=-3$, $\lambda_{2,3}=-3\pm\sqrt{9-3p\lambda^2}$. For $p\lambda^2>3$ this is a spiral critical point since the eigenvalues of the linearization matrix are complex numbers.

\end{itemize}

A brief summary of the main results can be given: i) noncommutative effects modify the early-times dynamics by erasing any past attractor (neither the stiff-fluid solution, nor the SF K/V-scaling solution can be source points), ii) noncommutative effects also modify the late-times dynamics: the stability of the SF potential energy-dominated solution (the future attractor in the standard GR-limit) is modified: it is now a saddle critical point. For $p\leq 1/2$ the future attractor in the phase space is the inflationary NC-dominated solution (equilibrium point $P_{NC}$). Additionally, the way in which the orbits in the phase space approach to the late-time attractor is also modified: while in the standard GR-limit the future attractor is a spiral point for $p\lambda^2>3/2$, in the general case, due to the influence of the noncommutative effects of the kind considered here, this attractor is a spiral point for $p\lambda^2>3$. The above enumerated features can be appreciated in the figure \ref{fig3}.

\section{Discussion}\label{discussion}

The question that stands in the title of this paper is not as trivial as it seems. To start with, noncommutativity, if it really have played any roll in the cosmological dynamics, is expected to have influenced the very early stages of the cosmic expansion, when, presumably, quantum effects had an appreciable impact on the gravitational interactions of matter. However, the relatively recent discovery that our universe expands at an accelerated peace, has put forward the possibility that tiny quantum effects that might have passed unnoticed in the recent past of the cosmic history, might be the cause of the present speed up of the expansion. Actually, the simplest model that accommodates the recent acceleration of the cosmic expansion rests on the anti-gravitating effect of the quantum vacuum (the cosmological constant). Since the energy density of the vacuum ($\rho_{vac}\propto\Lambda$) does not evolve with the expansion, then, even if it has been a tiny fraction of the matter-energy content of the universe in the past, as long as the remaining components of the cosmic mixture dilute with the expansion, there is a moment in the future (present) of the cosmic history when the negative pressure of the vacuum starts dominating the cosmic dynamics, resulting in a new period of inflation (also referred to as late-time inflation). 

Therefore, it makes sense to ask whether noncommutative effects can have any appreciable impact on the destiny of the cosmological evolution either. Besides, there is no evidence that its influence on the early-times dynamics is a generic feature. This is why we have focused in the study of the asymptotic properties of FRW cosmological models in connection with noncommutative quantum cosmology.

Here we have concentrated in a semi-classical WKB approximation to the minisuperspace WDW equation generalized to encompass noncommutativity of field variables $\alpha$ and $\vphi$ by the introduction of the Moyal star product. The resulting FRW cosmological equations (\ref{wkb-friedmann-eq}), (\ref{wkb-kg-eq}), (\ref{wkb-raycha-eq}) can be given an (attractive) alternative interpretation in the form of the equations (\ref{wkb-compact-eq}). Written in the latter form, the noncommutative effects of the kind considered in this paper, may be encoded in an additional (effective) NC matter term which equation of state tracks that of the cosmological constant $p_\theta=-\rho_\theta$, so that, its possible inflationary effect is envisioned. However, unlike the usual cosmological constant term, the energy density of the NC effective fluid does actually evolve as the cosmic expansion proceeds, thanks to additional non-gravitational interactions with the scalar field through a source term $\propto\dot\vphi\rho_\theta$.\footnote{Solar system measurements impose severe constraints to occurrence of additional non-gravitational interactions of matter, however, when these interactions involve dark components of the cosmic mixture, as in the present case, the issue is subtle and, in general, these constraints may be evaded \cite{ng-interactions,cqg2006}.} These interactions are switched on by the Moyal star product, which mixes the metric and the scalar field components.

Since, in general, the equations (\ref{wkb-friedmann-eq}), (\ref{wkb-kg-eq}), (\ref{wkb-raycha-eq}) (or their equivalent (\ref{wkb-compact-eq})) are very difficult to solve analytically, an alternative way around is to invoke the dynamical systems tools to extract very useful information about the asymptotic properties of the model instead. In this regard, knowledge of the equilibrium points in the phase space -- corresponding to a given cosmological model -- is a very important information since, independent on the initial conditions chosen, the orbits of the corresponding autonomous system of ODE will always evolve for some time in the neighborhood of these points. Besides, if the point were a stable attractor, independent of the initial conditions, the orbits will always be attracted towards it (either into the past or into the future). Going back to the original cosmological model, the existence of the equilibrium points can be correlated with generic cosmological solutions that might really decide the fate and/or the origin of the cosmic evolution. 

What the results of our dynamical systems analysis have revealed is that, independently of the kind of self-interaction potential considered: i) exponential potential, or ii) cosh-like potential, the noncommutative effects of the kind considered here\footnote{We frequently repeat this sentence because there are several different ways to build noncommutativity into a given cosmological setting.} affect not only the early-times dynamics but also modify the late-time behavior. Below we will discuss this issue in detail.

\subsection{Exponential potential}

Exponential potentials and their combination have been intensively studied in the recent past in connection with cosmological applications \cite{exp}. These arise in supergravity and in superstring after dimensional reduction \cite{exp-string}. 

The dynamical systems study of the NC cosmological model considered here reveals that, while for standard (GR) FRW cosmology the stiff fluid solution, $x=1$, is always the past attractor and the scaling solution, $x=\lambda/\sqrt 6\;\Rightarrow\;\dot\vphi^2/2V=\lambda^2/(6-\lambda^2)$, is the future attractor,\footnote{The case for $\lambda>\sqrt 6$, where the stiff fluid and the scaling solutions exchange their stability properties, is not being considered since this condition would imply negative energy.} in the most general case, which includes noncommutative effects, there are no past attractors at all. In other words, the noncommutative effects modify the early-time dynamics by erasing any past attractor from the phase space. This means, in turn, that the starting point of the cosmic dynamics is uncertain: if we evolve the cosmological equations from given initial data (given, say, in the present epoch) back into the past, the result is highly dependent on these data. However, the stiff fluid solution continues being an equilibrium point, which means that the cosmological evolution may evolve for some time in the vicinity of this solution. Hence, the noncommutative effects modify the stability properties of the equilibrium configuration associated with the early-times cosmic dynamics.

The surprising fact was to find that the NC effects also modify the late-times dynamics. Actually, in the general case, the SF kinetic/potential energy-scaling solution, $P_{K/V}=(\lambda/\sqrt 6,1)\;\Rightarrow\;\theta=0,\;\dot\vphi^2/2V=\lambda^2/(6-\lambda^2)$, is always a saddle critical point (it was the late-time attractor in the GR-limit), while the late-time attractor is the SF/NC fluid-scaling solution $P_{\vphi/\theta}=(\lambda/\sqrt 6,0)\;\Rightarrow\;\Omega_\vphi/\Omega_\theta=\lambda^2/(6-\lambda^2)$. This solution is inflationary (i. e., it is correlated with accelerated expansion of the universe) whenever $\lambda^2<2$.

\subsection{Cosh-like potential}

The cosh-like potential (\ref{cosh-pot}), $$V(\vphi)=V_0\left[\cosh(\lambda\vphi)-1\right]^p,$$ has been studied in connection with the so called scalar field dark matter (SFDM) models \cite{cosh,sahni,prd2009}. 

In a natural scenario for cosmic dynamics, the scalar field $\vphi$ runs from arbitrarily large negative values ($|\vphi|\gg 1$) to vanishing ones ($|\vphi|\ll 1$). In consequence at early times the dynamics is driven by an exponential potential $$|\vphi|\gg\frac{1}{\lambda}\;\Rightarrow\;V(\vphi)\approx\frac{V_0}{2}\;e^{-p\lambda\vphi},$$ whereas at late times it is associated with a power-law potential: $$|\vphi|\ll\frac{1}{\lambda}\;\Rightarrow\;V(\vphi)\approx\frac{1}{2}m^2\vphi^{2p},\;\;m^2\equiv V_0\lambda^2.$$

In general, for positive $p$-s, there is an oscillatory phase around the minimum of this potential which plays an important role in the late-time dynamics. It has been demonstrated \cite{turner} that in the limit when the oscillation period is much smaller than the time scale of the cosmic expansion, coherent scalar field (damped) oscillations behave like a fluid with $\left\langle p_\vphi\right\rangle=\left\langle \omega_\vphi\right\rangle\rho_\vphi$, where the mean equation of state of the fluid $\left\langle\omega_\vphi\right\rangle$ depends upon the form of the scalar field potential $V(\vphi)$ \cite{turner}. In particular, for $V=V_0\vphi^n\;\Rightarrow\;\left\langle\omega_\vphi\right\rangle=(n-2)/(n+2)$. Hence, for the potential (\ref{cosh-pot}), at late times -- during the oscillatory phase -- the mean equation of state is determined by $$\left\langle \omega_\vphi\right\rangle=\left\langle\frac{p_\vphi}{\rho_\vphi}\right\rangle=\frac{p-1}{p+1}.$$ Notice that for $p=1$ the scalar field behaves like pressureless dust, $\left\langle\omega_\vphi\right\rangle=0$. A scalar field potential with this value of $p$ could therefore play the role of cold dark matter (also known as SFCDM) in the universe. For, $p<1/2$, this potential is a good candidate for quintessence models of dark energy (DE) \cite{sahni,copeland}.

The known results of the dynamical systems study of this kind of potential -- within standard, commutative, FRW cosmology -- show that (see subsection \ref{cosh-gr}): i) The past attractor can be either the stiff fluid-dominated solution $P^+_K=(1,1/1+p\lambda)$ ($\omega_\vphi=1,\;\;q=2$), if $p\lambda<\sqrt 6$, or, whenever, $p\lambda>\sqrt 6$, it is the SF scaling solution $P_{K/V}=(p\lambda/\sqrt 6,1/1+p\lambda)$ ($\omega_\vphi=-1+p^2\lambda^2/3,\;\;q=-1+p^2\lambda^2/2$), ii) the SF potential energy-dominated solution $P_V=(0,1)$ ($\omega_\vphi=-1,\;\;q=-1$, meaning that $\vphi=\vphi_0$, $V(\vphi)=V_0$), is always the past attractor. For $p\lambda^2>3/2$ this is a spiral point signaling that the field $\vphi$ performs damped (coherent) oscillations around the minimum of the potential $V(\vphi)$ at late times, until the stable de Sitter solution is attained. This oscillatory stage is what can be identified properly with SFCDM \cite{prd2009}. The above features are clearly illustrated in the figure \ref{fig2}.

What kind of modifications of the above picture are produced by the noncommutative effects? According to the results of subsection \ref{cosh-general}, after NC effects of the kind explored here -- WKB approach to the minisuperspace WDW equation, supplemented with the Moyal star product -- are switched on, the past asymptotic structure of the phase space is modified: there are no past attractors there. 

The interesting finding is that the future asymptotics of the phase space is also modified. Actually, first, the stability of the SF potential energy-dominated solution (critical point $P_V$ above in subsection \ref{cosh-general}), which was the late-time attractor in the standard commutative theory, is modified: it is now a saddle equilibrium point. Second, an additional conjugated (twin) SF potential energy-dominated solution $P^*_V$ arises. Third, the oscillatory behavior of the perturbation is also modified by noncommutativity: while in the standard GR picture coherent oscillations arise for $p\lambda^2>3/2$, in the NC-modified picture the damped oscillations occur whenever $p\lambda^2>3$. Finally, the late-time attractor is the inflationary solution that is dominated by the NC effects (critical point $P_{NC}$ for which $\Omega_\theta=1$, see Fig. \ref{fig3}). The latter attractor exists whenever $p\leq 1/2$. Hence, the noncommutative effects may be the cause that the universe inflates at late times. 

A genuine objection against our finding that the present speed up of the cosmic expansion might be due to the effect of noncommutativity, can be based on the argument that the scalar field itself (alone) may fuel the late-time inflation. So that, why to complicate that simple quintessential picture with the inclusion of unknown noncommutative effects of quantum nature originated very early in the cosmic history? 

A reply to this kind of objections can be based on the following arguments. First, the SF models of dark energy are plagued with serious problems: fine tunning of initial conditions, the coincidence problem, etc. Besides the notion of dark energy itself is very unappealing and faces serious challenges to find support in the standard scheme of the fundamental interactions.\footnote{This is why alternative ways to explain the present stage of acceleration of the cosmic expansion have been explored. These include modifications of the laws of gravity. It is in this latter vein where our proposal fits. Modifications of gravity of quantum nature, such as inclusion of noncommutative effects, can be an interesting alternative to explain this -- up to date -- mysterious speed up of the expansion of the universe.}

A second argument is a bit more technical. Recall that a scalar field with a cosh-like potential of the kind (\ref{cosh-pot}), can be a nice model of dark matter if one chooses $p=1$ \cite{cosh,sahni,prd2009}, while, for $p<1/2$, it is a good candidate for quintessence model of dark energy instead \cite{sahni,copeland}. However, the latter statements are true only for non-interacting scalar fields -- no additional interaction with the other components of the cosmic mixture. In particular, in the present case where the scalar field interacts with the effective NC fluid, the above statements are not true. This can be clearly seen by recalling that the influence of noncommutativity can be alternatively understood as a modification to the SF self-interaction potential. Hence the range of values of the parameter $p$ for which the cosh-like potential correctly explains the dark matter, is shifted to lower values. Even if under the effects of noncommutativity, the (interesting for cosmology) picture with a late-time attractor inflationary solution, arises for $p\leq 1/2$, due to the mentioned shift in $p$, the scalar field with the potential (\ref{cosh-pot}) can be a good candidate to account for the dark matter. This statement is supported by the results of the dynamical systems study discussed above: for, $p\lambda^2>3$, coherent (damped) oscillations of the SF perturbation around the NC-dominated solution, arise. This oscillatory behavior is what can be interpreted as the (SF)CDM \cite{prd2009}. Hence, to close our line of argument; in our model the scalar field plays the role of CDM, while the noncommutative effects account for the late-time speed up of the expansion of the universe. 

In the Appendix, through the inspection of a simplified toy model, we avoid relying on the scalar field component to show that the noncommutative effects can, in fact, be a nice candidate to explain the cosmic speed up at late times.

\section{Conclusion}\label{conclusion}

In the present paper we have addressed the question, formerly explored in Ref. \cite{similar-dilaton}, about the possible impact of noncommutative effects of quantum nature on the dynamics at large cosmological scales, by the study of a simplified model \cite{prl-basic,citas-prl,citas-recientes-prl}. 

While in \cite{similar-dilaton} the issue was investigated in connection with a given particular solution of the corresponding modified cosmological equations (and for an exponential dilaton/scalar field self-interaction potential), here we have approached this subject from the dynamical systems perspective, and, additionally, have included the cosh-like potential which serves as a good model of cold dark matter, known as SFCDM \cite{cosh,sahni,prd2009}. 

The recipe used by us to build noncommutativity into the FRW (flat) cosmological model was based in the approach of reference \cite{prl-basic} (see also \cite{citas-prl,citas-recientes-prl}). It can be summarized in the following steps: i) the Hamiltonian is derived from the Einstein-Hilbert action (plus a self-interacting scalar field action) for a FRW spacetime with flat spatial sections, ii) canonical quantization recipe is applied, i. e., the minisuperspace variables are promoted to operators, and the WDW equation is written in terms of these variables, iii) noncommutativity in minisuperspace variables is achieved through the replacement of the standard product of functions by the Moyal star product in the WDW equation, and, finally, iv) (semi-classical) modified cosmological equations are obtained by means of the WKB approximation applied to the (equivalent) modified Hamilton-Jacobi equation.

Our results corroborate -- on the grounds of a solid dynamical systems basis -- the results of \cite{similar-dilaton} regarding the dynamics at late times. Noncommutativity does actually modify (appreciably indeed) the future asymptotics of the model: the late-time (future) attractor, whenever it exists, is associated with a solution of the cosmological equations that is dominated by the NC effects. Even in the case when the future attractor does not exist, the stability properties of the critical points associated with the future asymptotics are drastically modified. This is a robust result since it is not based on the study of a concrete particular solution, but on the exploration of the differential equations flux in the phase space corresponding to the original cosmological model.

However, in what regards the early-time dynamics, our results differ from the ones in reference \cite{similar-dilaton}. Actually, here we have shown that the past asymptotics is also modified by the noncommutativity: past attractors (also, source critical points) are erased by the NC effects. This is not an unexpected result since NC effects of quantum nature are designed to modify the cosmological evolution at early times, when, probably, the quantum effects played a major role in the gravitational dynamics. The discrepancy with the results of the mentioned reference may be a consequence of the different approaches undertaken. Besides, the fact that, according to the outcomes of \cite{similar-dilaton} the early-times dynamics is not affected by the noncommutativity, supports our argument that the solution studied therein was a particular, structurally unstable solution of the modified cosmological equations.

We want to recall that the present approach suffers from several ''drawbacks'': i) we have considered noncommutativity of minisuperspace variables rather than of spacetime coordinates themselves, and, ii) following the approach of \cite{prl-basic} we did not consider noncommutativity among the momenta conjugated of the minisuperspace variables. While the former drawback is not worrying since the kind of noncommutativity between the metric and the scalar field we have explored is expected to be a derived consequence of direct spacetime noncommutativity \cite{seiberg-witten,new-ref}, the latter drawback is of more concern. A more complete study along the lines followed here, where noncommutativity of the conjugated momenta is also considered, is the subject of ongoing research.

Since the modified cosmological model studied in the present paper can be, at most, a useful toy model to seek for qualitative aspects of the impact of noncommutativity at large scales, consequently we have not discussed any possible observational test to check it. Anyway, given that inclusion of noncommutativity the way it was included here, affects not only the late-time cosmic dynamics, but also the dynamics at very early times in a non-trivial way, it should be expected that the evolution of density perturbations in our model (in particular the grow of structure) is very different from the one predicted by other competing cosmological models, as for instance, the $\Lambda$CDM model, so that the new features can be detected by CMB measurements. The study of such an important issue deserves an independent publication. In the last instance, the present study can be considered as a first (modest) step towards a deeper understanding of the possible influence of primordial quantum processes in phenomena taking place at cosmological scales. 

The authors wish to thank Walberto Guzman and Miguel Sabido for most useful conversations. Due acknowledgments are given to PROMEP and CONACyT grant number 135023 by financial support of the present research, and also to Instituto Avanzado de Cosmologia (IAC) collaboration by partial support under grant number I0101/131/07 C-234/07.

\section{Appendix: Phenomenological toy model}\label{appendix}

Here we will speculate that, instead of the scalar field, one has a barotropic fluid with vanishing pressure $p_{DM}=0$ (dust), that can be identified with the cold dark matter in the universe. To do this in a consistent way within the framework of our NC model, it will suffice to notice that for a scalar field with vanishing pressure $p_\vphi=\dot\vphi^2/2-V=0\;\Rightarrow\;\rho_\vphi=\dot\vphi^2$. Then it is legitimate to make the replacement $\rho_\vphi\rightarrow\rho_{DM}\;\Rightarrow\;\dot\vphi\rightarrow\pm\sqrt{\rho_{DM}}$ (in what follows, for definiteness, we shall consider only the positive root in the latter expression). This procedure will result in the following set of cosmological equations for our NC phenomenological model (compare with equations (\ref{wkb-compact-eq})):

\bea &&3\dot\alpha^2=\rho_{DM}+\rho_\theta,\;\;2\ddot\alpha+3\dot\alpha^2=-p_\theta,\nonumber\\
&&\dot\rho_{DM}+3\dot\alpha\rho_{DM}=-\sqrt{\rho_{DM}}\rho_\theta,\;\;\dot\rho_\theta=\sqrt{\rho_{DM}}\rho_\theta.\label{toy-cosmo-eq}\eea 

The first thing we want to notice is that, as before, the cosmic dynamics is governed by the -- additional, non-gravitational -- interaction between the CDM and the NC fluid. A simple inspection of the above equations reveals that the pace of the expansion is determined by the correlation between the energy densities of both components of the cosmic mixture. Actually, it can be demonstrated that, in the present case, the deceleration parameter $q\equiv-(1+\ddot\alpha/\dot\alpha^2)$ can be expressed in the following form: $$q=-\frac{\rho_\theta-\rho_{DM}/2}{\rho_\theta+\rho_{DM}}.$$ 

Hence, if the CDM energy density dilutes with the cosmic expansion at higher rate than the NC fluid, as long as $\rho_\theta>\rho_{DM}/2$ the expansion transits from being decelerated in the past to being accelerated into the future. The above possibility, however, depends on the way $\rho_{DM}$ and $\rho_\theta$ evolve with the expansion of the universe. To make the discussion more precise it will be mandatory to go onto the phase space. The asymptotic structure will then clearly show which solutions are generic and, besides, will also reveal their stability properties.

To obtain an autonomous ODE out of the latter set of cosmological equations it will suffice to define a single phase space variable: $x\equiv\sqrt{\rho_\theta}/\sqrt 3\dot\alpha$. In terms of this variable the Friedmann constraint can be written in the following compact form: $\Omega_{DM}=\rho_{DM}/3\dot\alpha^2=1-x^2$. Since the dimensionless CDM energy density parameter has to be necessarily a non-negative quantity, then $|x|\leq 1$. However, since we are interested in cosmic expansion exclusively, negative values of $x$ will not be considered. Another useful quantity is the deceleration parameter $q=(1-3x^2)/2$. 

The following autonomous ordinary differential equation is obtained:

\be x'=\frac{\sqrt 3}{2}\;x\sqrt{1-x^2}[1+\sqrt{3(1-x^2)}].\label{aode-toy}\ee 

The critical values of the $x$-variable are: i) $x=0$ -- CDM-dominated ($\Omega_{DM}=1$), decelerated solution ($q=1/2$), and ii) $x=1$ -- NC fluid-dominated ($\Omega_\theta=1$), inflationary solution ($q=-1$). The CDM-dominated solution is unstable. Actually, let us perturb this solution, i. e., $x\rightarrow 0+\epsilon$. According to (\ref{aode-toy}) the perturbation $\epsilon$ will uncontrollably grow: $\epsilon(\alpha)=\epsilon(0)\exp((\sqrt 3+3)\alpha/2).$

To explore the stability of the NC-dominated solution it is recommendable to make the following replacement in equation (\ref{aode-toy}): $x\rightarrow y=\sqrt{1-x^2}$, so that the autonomous ODE can be written as: $2y'=-\sqrt 3(1-y^2)(1+\sqrt 3y).$ We recall that this equation is not valid at the point $y=0$ ($x=1$), since in the process of its derivation we divided by $y$. However, since we will be interested in perturbations around $y=0$, i. e., around $x=1$, but will not evaluate at the point $y=0$ itself, the above equation will be accurate enough. Lets now perform small perturbation $\epsilon$ around $y=0$. According to the latter equation the perturbation will decay as: $\epsilon(\alpha)\propto \exp(-3\alpha/2).$ In consequence the NC-dominated solution is stable. 

Stated in terms of the dynamical systems language: i) the CDM-dominated solution is the past attractor, while, ii) the NC-dominated solution is the future attractor. This demonstrates that the noncommutative effects alone can be, indeed, a candidate to explain the late-time speed up of the cosmic expansion.

The above is a nice cosmic scenario since transition from decelerated into accelerated expansion is generic. However, as with any toy model, it really does not correctly describes the past dynamics of the universe, since one needs to consider, also, other cosmic components as, for instance, a radiation term.

\end{document}